\documentclass[12pt,preprint]{aastex}

\shorttitle{Crystalline silicates around the Hen 3-600A}
\shortauthors{Honda et al.}

\begin{document}

\title{Detection of crystalline silicates around the T Tauri star Hen 3-600A\altaffilmark{1}}


\author{Mitsuhiko Honda\altaffilmark{2,3}, Hirokazu
Kataza\altaffilmark{4}, Yoshiko K. Okamoto\altaffilmark{5}, Takashi
Miyata\altaffilmark{6}, Takuya Yamashita\altaffilmark{3,2},
Shigeyuki Sako\altaffilmark{2,3}, Shinya Takubo\altaffilmark{2,3} and
Takashi Onaka\altaffilmark{2} 
}


\altaffiltext{1}{Based on data collected at Subaru Telescope, which is operated
by the National Astronomical Observatory of Japan.}
\altaffiltext{2}{Department of Astronomy, School of Science, University
of Tokyo, Bunkyo-ku, Tokyo 113-0033, Japan, onaka@astron.s.u-tokyo.ac.jp}
\altaffiltext{3}{Subaru Telescope, National Astronomical Observatory of
Japan, 650 North A'ohoku Place, Hilo, Hawaii 96720, U.S.A.,
hondamt@subaru.naoj.org, takuya@subaru.naoj.org, sako@subaru.naoj.org}
\altaffiltext{4}{Center for Advanced Spacecraft Technology, Institute of
Space and Astronautical Science, 3-1-1 Yoshinodai, Sagamihara, Kanagawa
229-8510, Japan, kataza@ir.isas.ac.jp}
\altaffiltext{5}{Institute of Physics, Center for Natural Science,
Kitasato University 1-15-1 Kitasato, Sagamihara, Kanagawa 229-8555,
Japan, okamtoys@cc.nao.ac.jp}
\altaffiltext{6}{Kiso Observatory, Institute of Astronomy, School of
Science, University of Tokyo, Mitake, Nagano 397-0101, Japan, 
miyata@kiso.ioa.s.u-tokyo.ac.jp}


\begin{abstract}
We have carried out mid-infrared N-band spectroscopic observations of
 the T Tauri star Hen 3-600A in the TW Hydra association with the COMICS on
 the 8.2m Subaru Telescope and found structured features in its spectrum. These
 structured features are well explained by a combination of crystalline 
forsterite, crystalline enstatite, silica and glassy
 olivine grains. 
 Among intermediate-mass young stellar objects (YSOs), crystalline silicates 
have already been detected, but no firm detection has been reported 
so far for low-mass YSOs such as T Tauri stars. This is the first clear
 detection of crystalline silicates in low-mass YSOs and shows
that the crystallization event occurs even in the protoplanetary disk of 
low-mass YSOs in the T Tauri phase. The physical processes leading to
 the inferred dust composition in the Hen3-600A system may be
analogous to those occured in the early epoch of the Solar system.
\end{abstract}





\section{Introduction}
Silicate is one of the main components of dust in the Universe. 
In the interstellar medium (ISM), it is believed to be completely amorphous
as implicated by the spectra with broad silicate absorption features
and no evidence for crystalline silicate has been indicated in the ISM.
Since the ISM material is the starting material for the star formation,
the presence of amorphous silicate in most YSOs is a natural consequence.
However, some Herbig Ae/Be stars \citep{hanner95,malfait98,sitko99,bouwman01} and 
the Vega-like star $\beta$Pic \citep{knacke93} exhibit the spectra which 
clearly deviates from the amorphous silicate features. 
Their spectra show crystalline silicate features and indicate the
presence of large grains,
which demonstrates dust processing in the protoplanetary disks.
By studying the dust composition of these YSOs, one can derive the physical 
processes occurring in, and the evolution of the circumstellar disks.
In addition, these YSO spectra resemble cometary spectra.
Comets often shows crystalline silicates \citep{hanner94,crovisier97},
preserving the record of the dust processing in the early phase of 
the proto solar nebula, suggesting that similar processes occurred in 
the early history of the solar system as are now observed in the YSOs.

However, most studies of dust processing so far made for YSOs are
limited to Herbig Ae/Be stars, which might be similar to, but are
obviously more massive than the solar-type stars.
It is thus important to investigate the dust processing in T Tauri
stars, which are supposed to be a good example of the early stage of the
solar-type stars, to understand the history of the solar system.
To date, among low-mass YSOs, there is no firm detection of crystalline silicates,
which is a typical indicater of dust processing.
Most mid-infrared spectroscopic studies of dust in low-mass YSOs
showed the amorphous silicate emission or absorption features \citep[e.g.][]{natta00}.
\cite{hanner95,hanner98} also concluded 
that the amorphous silicate emission accounts for their observed emission features 
of T Tauri stars and FU Ori stars and that the 11.2\,$\mu$m feature seen in some
objects can be attributed to polycyclic aromatic hydrocarbon (PAH) emission.
On the other hand, an unusually broad silicate emission feature was reported for HD98800
\citep[K5V;][]{skinner92,sylvester96}, a $\sim$10 Myr old pre-main 
sequence star (PMS) in the TW Hya Association (TWA). 
Such a broad silicate emission feature was thought to be the evidence for dust
growth \citep{skinner92,sylvester96}, while \cite{koener00} claimed that
crystalline silicates may account for it. 
TW Hya (K7Ve) is also a member of TWA, and is assumed to be an old classical 
T Tauri star (CTTS). Its mid-infrared spectrum was first presented by 
\cite{sitko00}. They could not detect the crystalline olivine feature at 
11.2 $\mu m$, and concluded that its silicate emission feature is similar to 
many other young pre-main sequence stars. However, \cite{weinberger02}
claimed hints for a crystalline olivine feature at 11.2$\mu $m in their N-band 
spectrum of TW Hya with a high S/N ratio and spectral resolution (R$\sim$120).

Since unusual silicate emission features were reported for old low-mass 
YSOs, such as TW Hya and HD98800, it would be a good strategy to investigate
old low-mass YSOs from the view point of silicate dust evolution. 
However, most old low-mass YSOs tend to be faint in the mid-infrared. 
Efforts to search for the evidence for crystalline silicates in
low-mass YSOs have so far been hampered by the faintness of the objects. 
The COoled Mid-Infrared Camera and Spectrometer (COMICS) on the
8.2m Subaru Telescope provides the higher sensitivity and enough spectral
resolution (R$\sim$250) to solve this issue. 
Therefore, we carried out the mid-infrared spectroscopic survey 
of low-mass YSOs, focusing on relatively old (5$\sim$10 Myr) objects.

In this {\it Letter}, we present the results of mid-infrared N-band
 spectroscopic observations of Hen3-600 from our mid-infrared
 spectroscopic survey with COMICS. It is a multiple system consisting of
 the spectroscopic binary primary star A (M3) and the secondary star B
 (M3.5), with a separation of 1.4$''$.
It is a member of the TWA, whose age is estimated to be 
1$\sim$10Myr old \citep[see discussions of ][]{jayawardhana99}.
\cite{muzerolle00} derived the mass accretion rate of Hen3-600A to be 5
$\times 10^{-11}$ M$_{\odot}$ yr$^{-1}$ from optical
 spectroscopy, which is by 1--2 orders of magnitude lower than the
 average rate in one-Myr old objects, suggesting the significant disk
 evolution and aging of the system.
Mid-infrared imaging observations by \cite{jayawardhana99} showed that
the the primary A has most of the circumstellar dust emission in this
system. They also pointed out a possibility of the inner hole in the
circumstellar disk from the SED of Hen3-600A. The present letter reports
 the clear detection of crystalline silicate features in old low-mass
 YSOs for the first time.

\section{Observations and Data Analysis}
We observed Hen 3-600 with the COoled Mid-Infrared
Camera and Spectrometer \citep[COMICS;][]{kataza00,okamoto02} 
mounted on the 8.2m SUBARU Telescope on Dec 27 2001. 
COMICS is an instrument for imaging
and long-slit spectroscopy in the 10 and 20\,$\mu$m atmospheric windows.
N-band low-resolution (R$\sim$250) spectroscopic observations were performed
with a 0.33'' wide slit that was set to run through Hen3-600A and B (The
position angle was $215^\circ$).
Imaging observations in the 8.8\,$\mu$m ($\Delta\lambda$=0.8$\mu$m) and 12.4$\mu
$m ($\Delta\lambda$=1.2$\mu$m) bands were also made for the calibration
of the
absolute flux of the spectra. To cancel the high background radiation,
the secondary mirror chopping was used at a frequency of 0.50Hz with a
10'' throw and a direction to the position angle of $215^\circ$. The
total on-source integration time was 598sec.
We selected HD123123 \citep{cohen99} as a rationing star for the
correction of the atmospheric absorption and a flux standard star for
aperture photometry. The observation parameters are summarized in Table \ref{tbl-1}.

\begin{figure}
\plotone{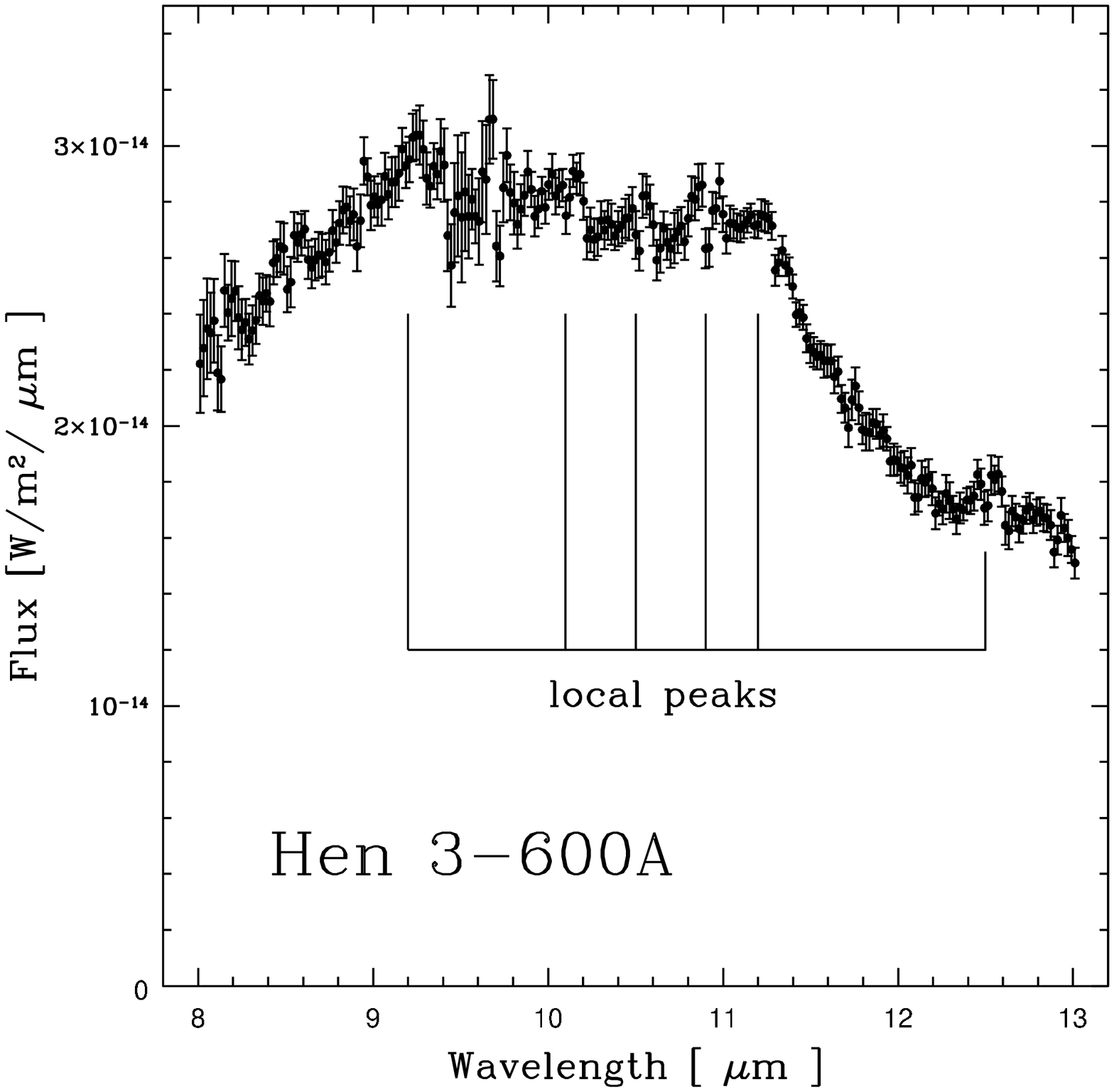}
\caption{Observed N-band spectrum of Hen3-600A. Local peaks at 9.2\,$\mu$m, 10.1\,$\mu$m, 10.5\,$\mu$m, 10.9\,$\mu$m, 11.2\,$\mu$m, and 12.5\,$\mu$m are indicated by straight lines.\label{fig1}}
\end{figure}

For the data reduction, we used our own reduction tools and IRAF.
For the imaging data, the standard reduction procedure and aperture photometry were applied.
For the spectroscopic data, the standard chopping subtraction and flat-fielding
by thermal spectra of the telescope cell-cover were made and the
distortion of the spectral image on the detector was corrected. Then the
spectra of Hen3-600A and of HD123123 were extracted. Due to the low chopping 
frequency, sky emission was not completely canceled out. Therefore we
derived the blank sky average level from the same object frame, 
and subtracted it as an offset from 
each extracted spectrum. The wavelength was calibrated by atmospheric
emission lines. 
After we divided the Hen3-600A spectrum by the HD123123
spectrum for the correction of atmospheric absorption, we multiplied the
resultant spectrum by the template of HD123123 \citep{cohen99}. Finally
we calibrated the absolute flux of the spectrum of Hen3-600A by the
photometric data in the 8.8\,$\mu$m and 12.4\,$\mu$m bands. 
The statistical uncertainty in the flux at each wavelength was
estimated from the noise in the blank sky and shown in Figure \ref{fig1}. 
The ozone absorption has increased the uncertainty in 9.3\, --
10.0\,$\mu$m due to the airmass mismatch between the objects and the
standards. We estimate the uncertainty at the wavelength of strongest ozone
absorption (9.6\,$\mu$m) to be at most 7\%.

\section{Results}
Figure \ref{fig1} shows the observed 8-13\,$\mu$m spectrum of Hen
3-600A. A broad silicate emission feature is clearly seen, whose profile
is much broader than that of amorphous silicate frequently seen for
a variety of astronomical objects. In addition,
it shows multiple local peaks located at 9.2\,$\mu$m, 10.1\,$\mu$m,
10.5\,$\mu$m, 10.9\,$\mu$m, 11.2\,$\mu$m and 12.5\,$\mu$m.

\begin{figure}
\plotone{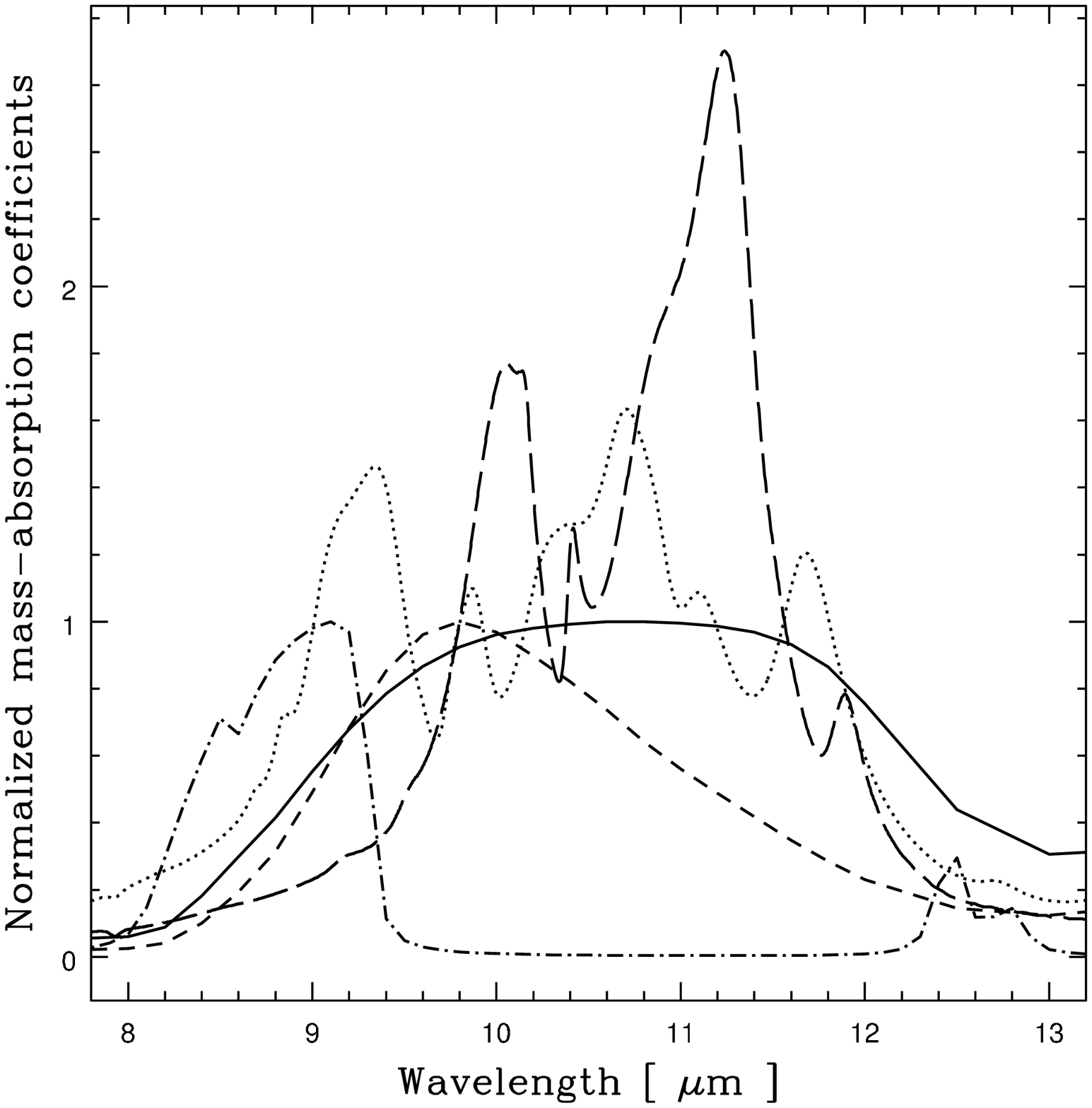}
\caption{Silicate dust spectral profiles of 0.1\,$\mu$m and 2.0\,$\mu$m glassy olivine \citep[dashed line, solid line;][]{dorschner95}, crystalline orthoenstatite \citep[dotted
 line;][]{koike00b}, crystalline forsterite \citep[long-dashed
 line;][]{koike00a}, and silica \citep[dot-dashed
 line;][]{spitzer61}. The normalizing constants in cm$^{2}$ g$^{-1}$ for
 the 0.1\,$\mu$m and 2.0\,$\mu$m glassy olivine, orthoenstatite, forsterite, and silica
 are 2372.3, 1664.6, 2451.9, 2476.6, and 16887.2, respectively.
\label{fig2}}
\end{figure}

To analyze it further, we made the following spectral model fitting.
We considered 4 dust species in our spectral model as described below.
The spectral profiles of these dust species are presented in Figure \ref{fig2}. \begin{enumerate}
\item Glassy olivine: For a representative of amorphous silicate, which is
often observed in the spectra of many YSOs, we used the optical
constants of glassy olivine by \cite{dorschner95}. Amorphous olivine
grains show little dependence on the grain shape, however, it depends on grain size \citep{bouwman01}.
To take grain size effect into account, we calculated the mass-absorption coefficient using the Mie theory for 0.1\,$\mu$m and 2.0\,$\mu$m spherical particles \citep{vandehulst57,matsumura86}.
The 0.1\,$\mu$m glassy olivine causes a narrow structure-less feature peaking at 9.8\,$\mu$m in the 10\,$\mu$m region, 
while that of the 2.0\,$\mu$m has a broad trapezoidal structure-less
feature (see Figure \ref{fig2}).
\item Crystalline forsterite: For the candidate for the observed structured 
features, we used the mass-absorption coefficient of crystalline forsterite
\citep{koike00a}. In the 8-13\,$\mu$m region, forsterite has emissivity peaks at 
10.06\,$\mu$m, 10.42\,$\mu$m, 11.24\,$\mu$m (strong), and 11.89\,$\mu$m.
\item Crystalline orthoenstatite: Pyroxene is one of the typical
      silicates as well as olivine. We used the mass-absorption
      coefficient of crystalline orthoenstatite \citep{koike00b}. 
Crystalline orthoenstatite of synthetic pure
      material has emissivity peaks at 9.34\,$\mu$m, 9.87\,$\mu$m,
      10.70\,$\mu$m, and 11.68\,$\mu$m in the 10\,$\mu$m region.
Both the optical data of forsterite and orthoenstatite
were obtained by the absorption measurements of small particles.  Their
shapes are rather random and may be represented by the continuous distribution
of ellipsoids (Koike et al., private communication).
\item Silica: Laboratory studies show that thermal 
annealing transforms the amorphous Mg silicate smokes into 
crystalline forsterite and crystalline/amorphous SiO$_2$ \citep{fabian00}. 
We take into account silica and use the parameters of \cite{spitzer61}. The optical
profile of silica is very sensitive to the grain shape. 
Small spherical particles can not account for the observed spectrum.
We have therefore calculated the mass-absorption coefficient by assuming a
continuous distribution of ellipsoids \citep[CDE; e.g.][]{bohren83} 
in the Rayleigh limit. The calculated mass-absorption coefficient has 
peaks at $\sim$9.1\,$\mu$m and $\sim$12.5\,$\mu$m.
\end{enumerate}
We did not consider PAH emission because PAH emission should be
associated with the features at 7.7\,$\mu$m, 8.6\,$\mu$m, 11.3\,$\mu$m, and
12.7\,$\mu$m, which are not detected for Hen3-600A.
The late spectral type (M3) of Hen3-600A is also consistent with the
absence of PAH emission because it cannot supply enough UV photons to
enhance PAH emission. Therefore, we consider only above 4 dust species
in the following analysis. 

\begin{figure}
\plotone{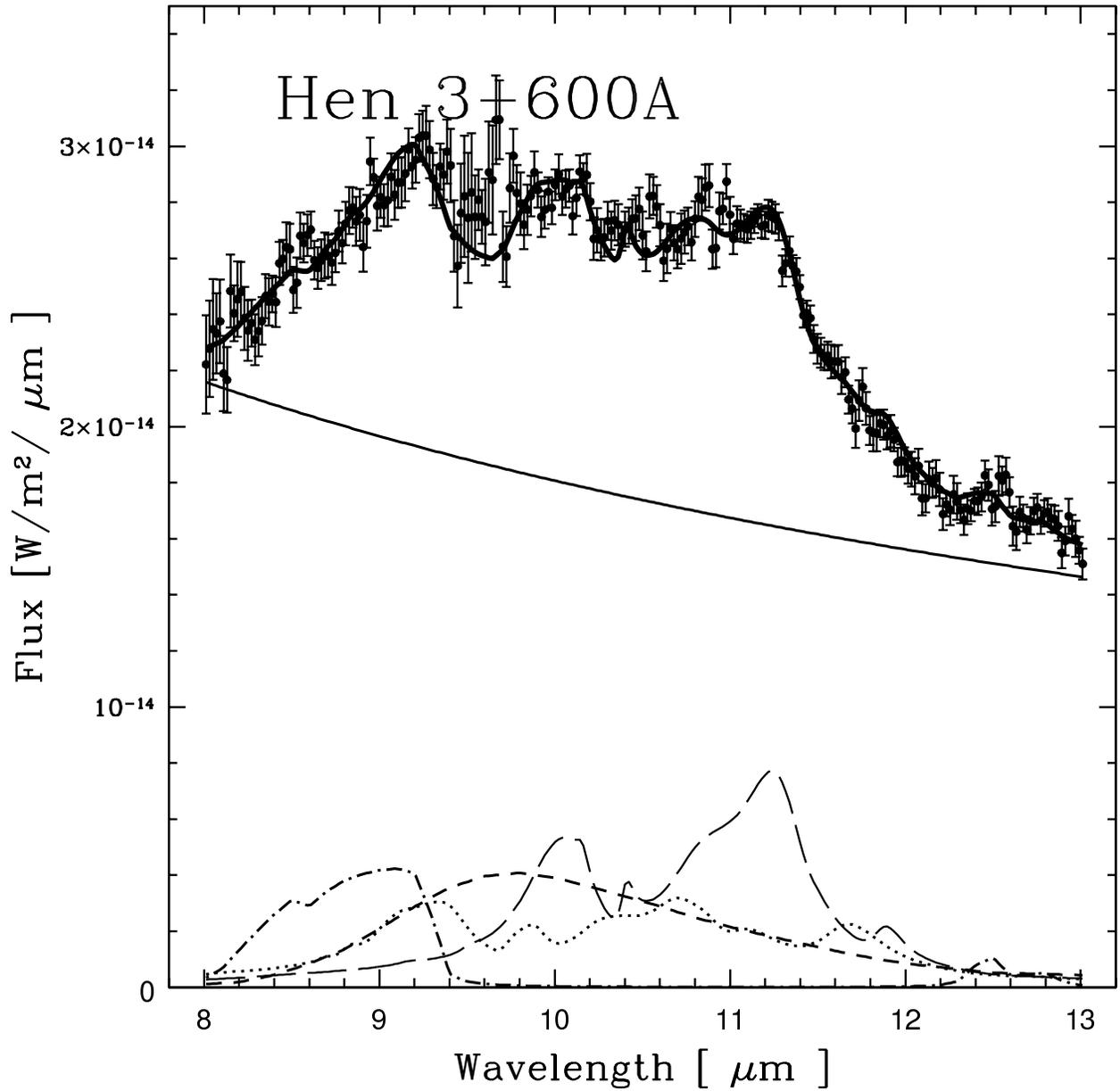}
\caption{Fit result of Hen3-600A. The thick solid line is the best-fit model
 spectrum, which is the sum of power-law continuum (thin solid), glassy
 olivine (dashed), orthoenstatite (dotted), forsterite (long-dashed), and silica (dot-dashed).
 \label{fig3}}
\end{figure}

In the model fit, we modified a case 2 model of
\cite{hanner95,hanner98}. The model flux $F_{\lambda}$ consists of two
components, featureless power-law continuum emission and optically thin
emission of the dust features. A power-law source function is
assumed. The model spectrum can be written by
\\$$\lambda F_{\lambda}=a_0(\frac{\lambda}{9.8\mu
m})^n+(a_1Q_{a0.1}+a_2Q_{a2.0}+a_3Q_{forst}+a_4Q_{ensta}+a_5Q_{silica})(\frac{\lambda}{9.8\mu m})^m,$$
where $Q_{a0.1},Q_{a2.0},Q_{forst},Q_{ensta},$ and $Q_{silica}$ are the absorption efficiencies of 0.1\,$\mu$m and 2.0\,$\mu$m glassy olivine, crystalline forsterite, crystalline
enstatite, and silica, respectively. They are shown in Figure \ref{fig2}. There are 8 free parameters
(6 multiplication factors $a_0,a_1,a_2,a_3,a_4$ in $10^{-14}$Wm$^{-2}$ and
2 spectral indices of source function, $n$ and $m$). 
We employed a least-squares minimization method to determine the most likely
parameters. Resulting best-fit parameters are $a_0$=1.8, $a_1$=0.4, $a_2$=0.0,
$a_3$=0.3, $a_4$=0.2, $a_5$=0.4, $n$=0.2, and $m$=0.5. The best-fit model spectrum using the derived parameters are shown in Figure \ref{fig3}.

Figure \ref{fig3} indicates that the 10\,$\mu$m spectrum of Hen 3-600A is
well reproduced by the model spectrum. The peaks at 10.1\,$\mu$m,
10.5\,$\mu$m, and 11.2\,$\mu$m are attributed to forsterite.
The 12.5\,$\mu$m and 9.2\,$\mu$m peaks can be owing to
 silica. The 10.9\,$\mu$m peak is likely to come from 
orthoenstatite, though the observed peak is slightly different from 
the orthoenstatite peak at 10.7\,$\mu$m. Other orthoenstatite features
are not seen in the spectrum or located at the wavelengths that are
disturbed by atmospheric ozone. Fe bearing pyroxene species can not account for the
peak shift \citep{chihara02}. However, orthoenstatite is needed to fit the
overall shape of model spectrum to the observed spectrum and as a candidate for 
the observed 10.9\,$\mu$m feature. 
The best-fit result indicates that no 2.0\,$\mu$m glassy olivine is required in 
the fitting and thus that the particle size effects do not account for the observed
band shape.
Based on these results, we conclude that a combination of glassy olivine, crystalline
forsterite, silica and possibly crystalline orthoenstatite accounts
for the observed spectrum. 

\section{Discussion}
At present no evidence for crystalline silicates in the ISM has been found,
and also the dust around many YSOs appears to be amorphous. 
Crystalline silicate features have so far been detected only in
the intermediate mass YSOs, Herbig Ae/Be stars 
\citep{hanner95,malfait98,sitko99,bouwman01}, while no firm detection of crystalline
silicates in T Tauri stars has been made to date.
Our detection of crystalline silicate features in the spectrum of the T
Tauri star Hen 3-600A indicates that the crystallization process does
occur even in the circumstellar disk of a low-mass YSO in the T Tauri phase.
Hen 3-600A is a TW Hya association member, and probably an old
CTTS evolving into a WTTS. Other TWA members, such as TW Hya and HD98800, are
also reported to show unusually broad silicate features, which indicate
silicate dust evolution \citep{skinner92,weinberger02}. Among low-mass YSOs,
silicate dust evolution is implicated only for objects in TWA. These
results may be attributed to the old age of TWA members, or they may also
be the characteristics specific to TWA. It is reported that
crystalline silicate grains tend to be present in old YSOs (old Herbig Ae/Be
stars) for intermediate-mass YSOs \citep{sitko99}. The present
observations suggest that this trend may also hold for low-mass YSOs. 
A possible simple explanation for this tendency
is that it takes a long time to produce an enough amount of crystalline
silicate grains to be observable. 
Alternatively, the crystallization process starts when a YSO gets old
enough. Further observational studies of a large sample are needed to
investigate this tendency.

From our fitting results of the observed features of Hen 3-600A, we
conclude that there are at least 4 dust species of glassy olivine,
crystalline forsterite, silica, and possibly crystalline enstatite. 
Crystalline forsterite is a major component in the crystalline
silicate observed to date. The presence of SiO$_2$ is expected based on
the laboratory thermal annealing experiment \citep{thompson01,thompson02,fabian00,hallenbeck00,rietmeijer02} and recent
studies of Herbig Ae/Be by \cite{bouwman01}.
The presence of enstatite in the YSO is a matter of debate. 
Its presence is rare in the YSO \citep{bouwman01}, though it is more commonly 
found around the evolved stars. In this paper, we attribute the 10.9\,$\mu$m 
feature of the Hen 3-600A spectrum to enstatite and 
it provides acceptable fitting results. However, \cite{thompson02} claimed that
the 10\,$\mu$m ``enstatite'' features are not sufficient evidence to confirm 
the presence of crystalline enstatite grains. 
While enstatite is so far the most favorite candidate to account for
the observed 10\,$\mu$m features, further investigations are clearly
needed to confirm the presence of enstatite in Hen 3-600A.
The derived mass ratio of these dust species (glassy olivine, forsterite,
enstatite and silica) is estimated to be 14:10:7:2,
assuming the same source function and mass distribution. 
Therefore crystalline silicate components consist of about 50\% 
(forsterite 30\%, enstatite 20\%) 
of the total small silicate grains in the optically thin emitting region. 
The derived mass ratio of forsterite and silica is consistent with 
the studies of Herbig Ae/Be by \cite{bouwman01}, which provided the constraint 
on the composition and structure of the amorphous material.

Low-mass YSOs are progenitors of our Solar system and crystalline
silicate grains are probably produced by thermal annealing
processes in the circumstellar disks. In the early
Solar system, there is evidence for grain heating, such as the formation
of chondrules found in primordial meteorites \citep{jones00}. Chondrules
are thought to be made by flush heating in the solar nebula.
There are several models for the formation of chondrules, such as
nebular lightening, shock heating, and X-wind models \citep{jones00}. Such
processes may also lead to annealing of amorphous silicate and prompt
silicate dust crystallization. The present detection of crystalline
silicates in the T Tauri star Hen3-600A suggests such heating process in
the early Solar system may be common among low-mass YSOs.

\acknowledgments

We are grateful to all of the staff members of SUBARU Telescope for
their support. We also thank Drs. Chiyoe Koike, Hiroki Chihara, Hiroshi Suto, 
and Koji Kawabata for providing us crystalline silicates spectra,
calculation codes, and useful comments. Finally, we would like to express thanks to
the referee Dr. J. Bouwman for helpful comments.

\clearpage

\clearpage 

\begin{deluxetable}{ccccc}
\tabletypesize{\scriptsize}
\tablecaption{COMICS Observations of Hen3-600\label{tbl-1}}
\tablewidth{0pt}
\tablehead{
\colhead{Mode} & \colhead{Object}   & \colhead{Filter [\,$\mu$m]}   &
\colhead{Integ. Time[sec]} & \colhead{Air Mass} 
}

\startdata
Imaging      &Hen 3-600 &8.8($\Delta\lambda =$0.8) &25.9 &1.879 \\
             &Hen 3-600 &12.4($\Delta\lambda =$1.2)&29.6 &1.855 \\
             &HD123123  &8.8($\Delta\lambda =$0.8) &5.1  &1.975 \\
             &HD123123  &12.4($\Delta\lambda =$1.2)&3.2  &1.754 \\
Spectroscopy &Hen 3-600 & -                        &598  &1.866-1.850\\
             &HD123123  & -                        &40.2  &1.769 \\ \enddata

\end{deluxetable}

\end{document}